\documentclass[RNAAS]{aastex63}

\graphicspath{{./}}
\begin{document}

\title{TESS Science Processing Operations Center Photometric Precision
Archival Product}

\correspondingauthor{Joseph D. Twicken}
\email{joseph.twicken@nasa.gov}

\author[0000-0002-6778-7552]{Joseph D. Twicken}
\affiliation{SETI Institute, Mountain View, CA 94043, USA}
\affiliation{NASA Ames Research Center, Moffett Field, CA 94035, USA}

\author[0000-0002-4715-9460]{Jon M. Jenkins}
\affiliation{NASA Ames Research Center, Moffett Field, CA 94035, USA}

\author[0000-0003-1963-9616]{Douglas A. Caldwell}
\affiliation{SETI Institute, Mountain View, CA 94043, USA}
\affiliation{NASA Ames Research Center, Moffett Field, CA 94035, USA}

\author[0000-0003-2053-0749]{Benjamin M. Tofflemire}
\affiliation{SETI Institute, Mountain View, CA 94043, USA}
\affiliation{NASA Ames Research Center, Moffett Field, CA 94035, USA}

\author[0000-0001-8019-6661]{Marziye Jafariyazani}
\affiliation{SETI Institute, Mountain View, CA 94043, USA}
\affiliation{NASA Ames Research Center, Moffett Field, CA 94035, USA}

\author[0000-0002-1949-4720]{Peter Tenenbaum}
\affiliation{SETI Institute, Mountain View, CA 94043, USA}
\affiliation{NASA Ames Research Center, Moffett Field, CA 94035, USA}

\author[0000-0002-6148-7903]{Jeffrey C. Smith}
\affiliation{SETI Institute, Mountain View, CA 94043, USA}
\affiliation{NASA Ames Research Center, Moffett Field, CA 94035, USA}

\author[0009-0008-5145-0446]{Stephanie L. Striegel}
\affiliation{SETI Institute, Mountain View, CA 94043, USA}
\affiliation{NASA Ames Research Center, Moffett Field, CA 94035, USA}

\author[0000-0002-8219-9505]{Eric Ting}
\affiliation{NASA Ames Research Center, Moffett Field, CA 94035, USA}

\author[0000-0002-5402-9613]{Bill Wohler}
\affiliation{SETI Institute, Mountain View, CA 94043, USA}
\affiliation{NASA Ames Research Center, Moffett Field, CA 94035, USA}

\author[0000-0003-4724-745X]{Mark E. Rose}
\affiliation{NASA Ames Research Center, Moffett Field, CA 94035, USA}

\author[0000-0003-2196-6675]{David Rapetti}
\affiliation{Research Institute for Advanced Computer Science, Universities Space Research Association, Washington, DC 20024, USA}
\affiliation{NASA Ames Research Center, Moffett Field, CA 94035, USA}

\author[0000-0002-9113-7162]{Michael M. Fausnaugh}
\affiliation{Department of Physics and Astronomy, Texas Tech University, Lubbock, TX 79409, USA}

\author[0000-0001-6763-6562]{Roland Vanderspek}
\affiliation{Kavli Institute for Astrophysics and Space Research, Massachusetts Institute of Technology, Cambridge, MA 02139, USA}

\begin{abstract}
We report the delivery to the Mikulski Archive for Space Telescopes (MAST) of tables containing Root Mean Square (RMS) Combined Differential Photometric Precision (CDPP) values for all TESS 2-min cadence targets with Science Processing Operations Center (SPOC) light curves in Sectors 1–90. Each comma-separated values (CSV) file contains CDPP values for all 2-min light curves in the given sector. The tables include robust RMS CDPP values for the 15 trial transit pulse durations searched in the SPOC 2-min processing pipeline, ranging from 0.5–15.0 hr. For each pulse duration, CDPP is computed in the transit search for a trial transit centered on every cadence. The RMS value of the CDPP time series is a metric that may be employed to estimate signal-to-noise ratio for transits with the given duration and a specified depth. We will continue to deliver the RMS CDPP tables to MAST for each observing sector.
\end{abstract}



\section{} 

Since launch in 2018, the TESS mission \citep{rickerTESS} has observed most of the sky in search of transiting exoplanets. With four cameras, a 24$^\circ$ x 96$^\circ$ region of the sky is covered in $\sim$28~day observing sectors. The TESS Science Processing Operations Center \citep[SPOC;][]{SPOC} pipeline has performed photometry and conducted a transit search of systematic error corrected light curves for up to 20,000 targets observed at 2-min cadence in each sector. The SPOC periodically conducts searches of multiple-sector light curves to improve sensitivity to small planets and those with relatively long orbital periods. The SPOC also operates a functionally equivalent pipeline for Full Frame Image (FFI) based target stars \citep{FTL} and delivers archival products to the Mikulski Archive for Space Telescopes\footnote{\url{https://archive.stsci.edu/}} (MAST) as TESS-SPOC High Level Science Products (HLSP).

The SPOC pipeline transit search employs an adaptive, noise-compensating matched filter \citep{jenkins2002tps, jenkins2010tps, jenkins2020tps}. The noise component for characterizing transit signal-to-noise ratio (S/N) is the photometric precision on the time scale of the transit. Photometric precision is measured for each trial pulse duration in the form of Combined Differential Photometric Precision \citep[CDPP;][]{jenkins2010tps, christiansen2012}. The trial transit pulse durations (i.e., integration times) range from 0.5--15.0~hr in the 2-min pipeline. For each pulse duration, CDPP is computed for all 2-min cadences (after filling data gaps with an algorithm intended to preserve statistical properties). The CDPP time series reflect the time-varying noise characteristics associated with the SPOC Presearch Data Conditioning Simple Aperture Photometry light curve \citep[PDCSAP; ][]{stumpe2012, stumpe2014, smith2012} associated with each target star.

The robust Root Mean Square (RMS) value of a CDPP time series represents a scalar photometric precision metric for the given integration time. The robust RMS value of CDPP time series, $X$, is calculated in the pipeline by:
\begin{equation}
X_{\mathrm{robust-rms}} = \sqrt{(X_{\mathrm{median}})^2 + (1.4826 \cdot X_{\mathrm{mad}})^2}
\end{equation}
where $X_{\mathrm{median}}$ is the median value, and $X_{\mathrm{mad}}$ is the median absolute deviation of the time series (scaled here as an estimator of standard deviation for Gaussian noise). Robust RMS CDPP values are provided in headers of the SPOC Target Pixel and Light Curve files for integration times of 0.5, 1.0, and 2.0~hr with keywords \texttt{CDPP0\_5}, \texttt{CDPP1\_0}, and \texttt{CDPP2\_0} respectively \citep{twicken2020}. Target Pixel and Light Curve files are delivered to MAST for each 2-min target star and sector in which the target was observed.\footnote{It should be noted that SPOC does not produce Light Curve files for a handful of target stars in each observing sector, typically because they exceed a brightness threshold. These targets are enumerated in the Data Release Notes posted at MAST for each observing sector.}

\begin{figure}[htb]
\begin{center}
\includegraphics[scale=0.19,angle=0]{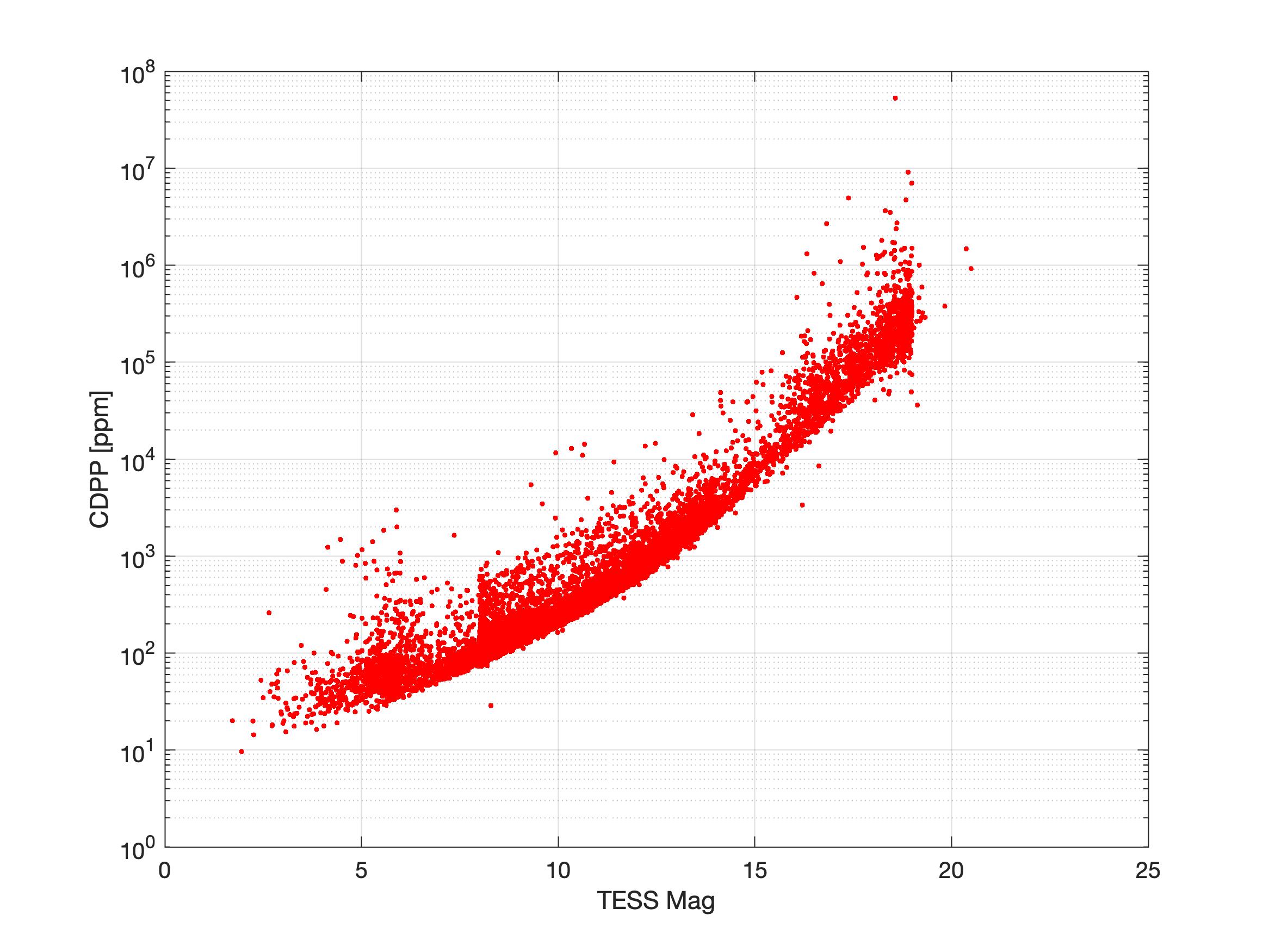}
\caption{Robust RMS CDPP values at 1~hr integration time displayed as a function of TESS magnitude for 12,995 SPOC 2-min light curves in Sector~89. CDPP is shown on a logarithmic scale.\label{fig:1}}
\end{center}
\end{figure}

We report here on the delivery to MAST of comprehensive robust RMS CDPP files in comma-separated values (CSV) format for all 2-min targets with SPOC light curves in each of Sectors 1--90.\footnote{The RMS CDPP files are archived at MAST: \href{https://doi.org/10.17909/xx44-3n34}{doi:10.17909/xx44-3n34}.} Hereafter, the CDPP files will be delivered as each sector processing is completed. The CDPP table columns include the following:
\begin{itemize}
    \item \texttt{ticid} -- TESS Input Catalog \citep[TIC;][]{stassunCTL} target star identifier.
    \item \texttt{tmag} -- TESS magnitude \citep{stassunCTL} from latest TIC version at time of transit search.
    \item \texttt{rrmscdppMMpN} (15x) -- Robust RMS CDPP in parts per million (ppm) at integration time \textit{MM.N} in hr; for 2-min cadence, the integration times are 0.5, 1.0, 1.5, 2.0, 2.5, 3.0, 3.5, 4.5, 5.0, 6.0, 7.5, 9.0, 10.5, 12.5, and 15.0 hr.
\end{itemize}
The value of robust RMS CDPP is archived as NaN (i.e.,~Not a Number) for target stars and integration times where RMS CDPP could not be computed. For sectors where SPOC has reprocessed data to produce new archival Target Pixel and Light Curve products, the files in this delivery include RMS CDPP values from the latest pipeline processing. The robust RMS CDPP values at 1~hr for all SPOC 2-min light curves in Sector 89 are shown in Figure~\ref{fig:1}.

The CDPP file naming convention follows the TESS Science Data Products Description Document standard \citep{twicken2020}: `tess\textit{yyyydddhhmmss}-s\textit{startsctr}-s\textit{endsctr}-\textit{pin}\_rms-cdpp.csv' where \textit{yyyydddhhmmss} encodes a sector start timestamp, \textit{startsctr} and \textit{endsctr} are 4-digit zero-padded values indicating the first and last sectors associated with the given transit search, and \textit{pin} is a 5-digit zero-padded pipeline instance identifier indicating the specific transit search that produced the CDPP values. The naming convention allows for the possibility of archiving RMS CDPP files for multiple-sector SPOC transit searches. The naming of CDPP files associated with FFI-based transit searches is likely to differ because MAST maintains an alternative convention for HLSP deliveries. If CDPP files are produced for light curves extracted from FFIs, the pulse durations for the RMS CDPP values may not identically match the integration times for 2-min targets quoted earlier.

The archival RMS CDPP values facilitate the estimation of transit S/N in searches of SPOC PDCSAP light curves. For a light curve with $N$ transits of depth $\delta$ (in ppm) and duration $\tau$ (in hr) in a given sector, detection S/N (denoted $S$) may be estimated as follows:
\begin{equation}
S = \frac{\delta}{C(\tau)} \cdot \sqrt{N}
\label{eq:snn}
\end{equation}
where $C(\tau)$ represents RMS CDPP at an integration time close to $\tau$ in the same sector as the transits. The implicit assumption is that the transit events are independent. This is reasonable given that the orbital period defining the spacing between transits is typically long in comparison with the transit duration. In the case of $N$ transits observed across multiple sectors such that $N_{i}$ transits are observed in sector $i$, the estimated S/N may be generalized by:
\begin{equation}
S = \frac{\delta}{\sqrt{\sum_{i} N_{i}\, C_{i}^{2}(\tau)\, /\, N}} \cdot \sqrt{N} =
\frac{\delta}{\sqrt{\sum_{i} N_{i}\, C_{i}^{2}(\tau)}} \cdot N
\label{eq:snnms}
\end{equation}
where $C_{i}(\tau)$ represents RMS CDPP at an integration time near $\tau$ in sector $i$.


\acknowledgments

This paper is based on data collected by the TESS mission. Funding for the TESS mission is provided by the NASA Explorer Program. Resources were provided by the NASA High-End Computing (HEC) Program through the NASA Advanced Supercomputing (NAS) Division at Ames Research Center for the production of SPOC data products and TESS-SPOC HLSP.

\facilities{\textit{TESS}}

\bibliography{tess-cdpp}{}
\bibliographystyle{aasjournal}

\end{document}